\documentclass[12pt,noshowpacs,nofootinbib,notitlepage,amsmath]{revtex4-1}
\allowdisplaybreaks
\usepackage{graphicx,color}
\usepackage[colorlinks=true,citecolor=blue,linkcolor=blue,urlcolor=blue]{hyperref}
\usepackage[charter]{mathdesign}
\DeclareSymbolFontAlphabet{\mathcal}{symbols}
\DeclareSymbolFont{symbols}{OMS}{xmdcmsy}{m}{n}
\DeclareSymbolFont{largesymbols}{OMX}{xmdcmex}{m}{n}
\SetSymbolFont{symbols}{bold}{OMS}{xmdcmsy}{b}{n}

\begin{document} 
\title{\color{blue}\Large Focus Point  in Dark Matter Selected High-Scale Supersymmetry}
\author{Sibo Zheng}
\email{sibozheng.zju@gmail.com}
\affiliation{ Department of Physics, Chongqing University, Chongqing 401331, P. R. China}
\begin{abstract}
In this paper, 
we explore conditions for focus point in the high-scale supersymmetry 
with the weak-scale gaugino masses.
In this context the tension between the naturalness and LHC 2013 data about supersymmetry 
as well as the cold dark matter candidate are addressed simultaneously.
It is shown that the observed Higgs mass can be satisfied in a wide classes of new models,
which are realized by employing the non-minimal gauge mediation.
\end{abstract}
\maketitle

\section{Introduction}
The standard model (SM)-like Higgs scalar with mass around 125 GeV \cite{125} 
discovered at the LHC needs some mechanism for stabilizing it against high-energy scale quantum correction.
Among the well known candidates which achieve this naturally
low-scale supersymmetry (SUSY) is expected to play an important role at the TeV scale. 
However,  the first run of LHC has not observed any signal of new physics yet,
and pushes the SUSY particle masses into multi-TeVs region \cite{ATLAS2013, CMS2013}.
So the absence of SUSY particles near the weak scale $v$, 
together with the observed Higgs mass, 
severely challenge the low-scale SUSY.

In the context of minimal supersymmetric standard model (MSSM)
stop masses far above the weak scale is required by the observed Higgs mass 
when the mixing effect is weak.
Given SUSY mass spectrum far above the weak scale, 
the naturalness is spoiled naively.
However, this statement can be relaxed in some specific situations.
SUSY with focusing phenomenon \cite{9908309,9909334}, 
which is named as focus point SUSY, is few of such examples.
In focus point SUSY the sensitivity of up-type Higgs mass squared to the mass scale of SUSY mass spectrum  
is suppressed because of cancellation among the large renormalization group (RG) corrections.
This phenomenon leads to a dramatical reduction of fine tuning associated with electroweak symmetry breaking (EWSB). 
As a result, it provides us an alternative choice of natural SUSY 
consistent with the LHC data and the observed Higgs mass.

Unfortunately, focus point SUSY can not be realized in the minimal setup 
from the viewpoint of model building such as conventional supergravity \cite{1205.2372} 
and the minimal gauge mediation (GM) \cite{9910497}.
However, they are expected to be achieved in some subtle cases.
For recent examples in the context of GM, see, e.g., \cite{1312.4105,1312.5407,1402.1735,1403.6527}.
These examples are restricted to special SUSY-breaking mediation scale $M$.

Based on our earlier work \cite{1312.4105},  
we will take an arbitrary $M$ and generalize results above to general focus point.
Consider the fact that the cold dark matter demands some electroweakino mass around the weak scale,
we will focus on the high-scale SUSY 
in which gaugino masses are light in compared with other SUSY particle masses \footnote{
For recent discussion about the prediction of Higgs mass in the high-scale SUSY 
with dark matter mass of the weak scale, see, e.g., \cite{1409.2939} and references therein.}.

The paper is organized as follows.
Firstly, we determine the conditions for focus point in SUSY with arbitrary $M$ in section II.
Then, we analyze the prediction for the Higgs mass in focus point SUSY in section III.
We find that the observed Higgs mass requires the input value for $m^{2}_{H_{u}}$ of order $\sim$ multi-TeVs.
With such magnitude of $m^{2}_{H_{u}}$ the observed Higgs mass can be explained in a wide classes of high-scale SUSY models.
The second part of this paper is devoted to the realization of focus point in high-scale SUSY.
In section IV we will construct concrete and complete examples by employing non-minimal GM.
We will show that for the case of small $A_t$ term 
focus point SUSY is viable for a wide range of $M$,
which generalizes previous results in the literature.
For the case of large $A_t$ term, 
we find that focus point SUSY is viable 
in a large classes of GM with direct Yukawa coupling 
between the messengers and the MSSM singlet.

Finally we discuss our results in section V.
The calculation of soft masses is included in appendix A.

\section{Conditions for Focus Point}
We begin with the conditions for focus point in SUSY.
Given light gauginos, in compared with soft masses 
$m^{2}_{\tilde{Q}_{3}}$, $m^{2}_{\tilde{U}_{3}}$, $m^{2}_{H_{u}}$ and  $A_t$ term squared, 
the one-loop renormalization group equations (RGEs) for them are simply given by, 
\begin{eqnarray}{\label{RGE}}
\frac{\partial}{\partial \ln t} m^{2}_{H_{u}} &=& 3\cdot \frac{y^{2}_{t}}{8\pi}\left(m^{2}_{\tilde{Q}_{3}}+ m^{2}_{\tilde{U}_{3}} +m^{2}_{H_{u}}+A^{2}_{t}\right),\nonumber\\
\frac{\partial}{\partial \ln t} m^{2}_{\tilde{U}_{3}} &=& 2\cdot \frac{y^{2}_{t}}{8\pi}\left(m^{2}_{\tilde{Q}_{3}}+ m^{2}_{\tilde{U}_{3}} +m^{2}_{H_{u}}+A^{2}_{t}\right),\nonumber\\
\frac{\partial}{\partial \ln t} m^{2}_{\tilde{Q}_{3}} &=&1\cdot \frac{y^{2}_{t}}{8\pi}\left(m^{2}_{\tilde{Q}_{3}}+ m^{2}_{\tilde{U}_{3}} +m^{2}_{H_{u}}+A^{2}_{t}\right),\nonumber\\
\frac{\partial}{\partial \ln t} A^{2}_{t} &=& 12\cdot \frac{y^{2}_{t}}{8\pi}A^{2}_{t},
\end{eqnarray}
where $t\equiv Q/M$, with $Q$ the RG running scale and 
$M$ the  SUSY-breaking mediation scale.
Here $y_{t}$ denotes the top Yukawa coupling. 
All other SM Yukawa couplings will be ignored in our discussion.
The correlation for these soft masses between scale $M$ and the weak scale $v=174$ GeV is obtained in terms of solving the RGEs in  Eq.\eqref{RGE}.
In particular, we have
\begin{eqnarray}{\label{mH}}
2m^{2}_{H_{u}}(v)&=&I[M]\left(m^{2}_{\tilde{Q}_{3}}[M]+ m^{2}_{\tilde{U}_{3}}[M] +m^{2}_{H_{u}}[M]-A^{2}_{t}[M]\right)\nonumber\\
&+&I[M]^{2}A^{2}_{t}[M]-\left(m^{2}_{\tilde{Q}_{3}}[M]+ m^{2}_{\tilde{U}_{3}}[M]-m^{2}_{H_{u}}[M]\right),
\end{eqnarray}
where the real coefficient $I[M]$ is defined as,
\begin{eqnarray}{\label{I}}
I[M]=\exp\left( 6\times \int^{\ln\frac{v}{M}}\frac{ y_{t}^{2}[t']}{8\pi^{2}} dt' \right).
\end{eqnarray}
For the case with negligble $A_{t}$, Eq.\eqref{mH} reduces to the well known result \cite{9908309, 9910497},
\begin{eqnarray}{\label{noA}}
2m^{2}_{H_{u}}(v)&=&I[M]\left(m^{2}_{\tilde{Q}_{3}}[M]+ m^{2}_{\tilde{U}_{3}}[M] +m^{2}_{H_{u}}[M]\right)-\left(m^{2}_{\tilde{Q}_{3}}[M]+ m^{2}_{\tilde{U}_{3}}[M]-m^{2}_{H_{u}}[M]\right).
\end{eqnarray}

As well known, for soft mass spectrum far above the weak scale in Eq.\eqref{mH} there is a large fine tuning naively.
But this can be avoided if there is significant cancellation among large contributions 
to up-Higgs soft mass squared in Eq.\eqref{mH}.
Generally speaking,  this cancellation does not happen in the minimal setup of model building.
However, if it indeed happens in such as focus point SUSY, 
with the cancellation referred as focusing phenomenon, 
one can derive the focus condition, regarding Eq.\eqref{mH}.

\linespread{1}\begin{figure}[!h]
 \centering%
\begin{minipage}[b]{0.75\textwidth}
\centering
\includegraphics[width=4.5in]{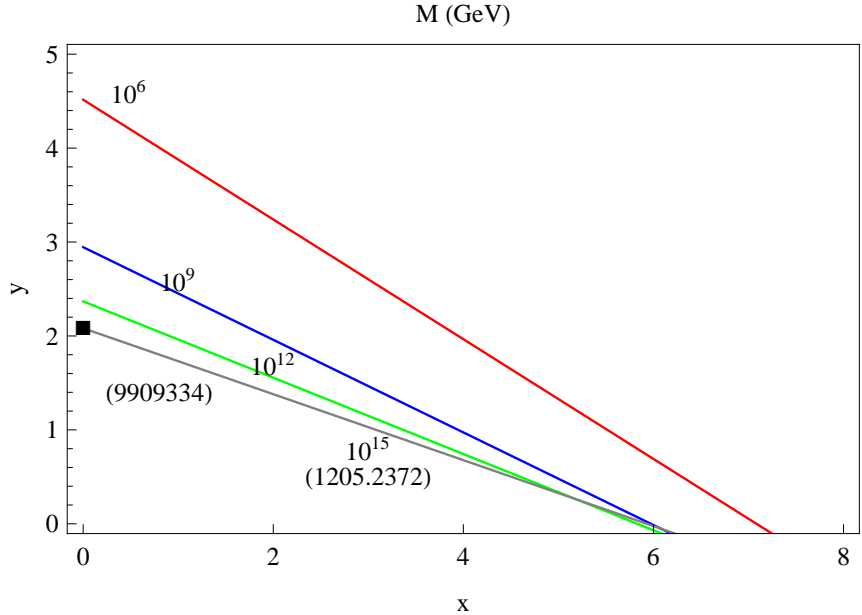}
\end{minipage}%
\caption{Focusing lines in the two-parameter plane of $(x, y)$  for $M=\{10^{6}, 10^{9}, 10^{12}, 10^{15}\}$ GeV. Any point in each focusing line generates the focusing phenomenon, $m^{2}_{H_{u}}[v]\simeq 0$. See the text for comments on references.}
\label{focuscondition}
\end{figure}

Contrary to the case with small $A_t$ term, where only two real parameters are needed,
we define $m^{2}_{H_{u}}[M]\equiv m^{2}_{0}$ and introduce three dimensionless real parameters $x$, $y$ and $z$,
\begin{eqnarray}{\label{definition}}
x \equiv \frac{A^{2}_{t}[M]}{m^{2}_{H_{u}}[M]},~~~~~~
y \equiv \frac{m^{2}_{\tilde{Q}_{3}}[M]+ m^{2}_{\tilde{U}_{3}}[M]}{m^{2}_{H_{u}}[M]},~~~~~~
z\equiv \frac{m^{2}_{\tilde{Q}_{3}}[M]}{m^{2}_{H_{u}}[M]}.
\end{eqnarray}
In terms of Eq.\eqref{definition} soft masses at scale $M$ can be simply parameterized by $x$, $y$ and $z$, 
\begin{eqnarray}{\label{correlation}}
\left(%
\begin{array}{c}
  m^{2}_{H_{\mu}} [M]\\
  m^{2}_{\tilde{Q}_{3}} [M]\\
  m^{2}_{\tilde{U}_{3}} [M]\\
  A^{2}_{t} [M] \\
\end{array}%
\right)=
m^{2}_{0}\left(%
\begin{array}{c}
  1\\
  z \\ 
  y-z\\
  x\\
\end{array}%
\right)
\rightarrow
\left(%
\begin{array}{c}
  m^{2}_{H_{\mu}} [v]\\
  m^{2}_{\tilde{Q}_{3}} [v]\\
  m^{2}_{\tilde{U}_{3}} [v]\\
  A^{2}_{t} [v]) \\
\end{array}%
\right)=
m^{2}_{0}\left(%
\begin{array}{c}
  0\\
  z-\frac{1}{3}\\ 
  y-z-\frac{2}{3}\\
  I^{2}[M] x\\
\end{array}%
\right),
\end{eqnarray}
and their values at the weak scale are obtained through the RGEs in Eq.\eqref{RGE}.
Note that we have imposed the ``focusing '' condition $m^{2}_{H_{u}}[v]=0$,
which leads to a constraint on the input masses from Eq.\eqref{mH},
\begin{eqnarray}{\label{condition}}
-x\left(I^{2}[M]-I[M]\right)+y\left(1-I[M]\right)-1-I[M]=0.
\end{eqnarray}
Moreover, from Eq.\eqref{correlation} one reads the allowed ranges for the parameters as,
\begin{eqnarray}{\label{range}}
x\geq 0 ~~~~\text{and}~~~~~ \frac{1}{3}<z< y-\frac{2}{3}.
\end{eqnarray}

In Fig.\ref{focuscondition} we show the ``focusing '' line in the plane of $x$ and $y$ for $M=\{10^{6}, 10^{9}, 10^{12}, 10^{15}\}$ GeV.
Each point on the line gives rise to the focusing phenomenon. 
Earlier works in \cite{9909334, 1312.5407} discussed focusing in GUT-scale model without $A$ term.
In these cases $I[M_{\text{GUT}}]\simeq 1/3$, so
they correspond to the focus point $(0, 2)$ in Fig.\ref{focuscondition}.
Work in \cite{1205.2372} addressed similar situation but with sizable $A$ term,
it corresponds to the gray line.
Work in \cite{1312.4105} discussed intermediate scale model with $M\sim 10^{8}$ GeV,
which corresponds to the blue line.

\section{Higgs Mass in Focus Point SUSY}
In this section we discuss the prediction for the Higgs mass in high-scale SUSY with focus point.
In particular, we would like to show which points in the focus lines in Fig.\ref{focuscondition} 
can explain the observed Higgs mass $m_{H}=125. 5$ GeV \cite{125} at the LHC. 
It is well known that the tree-level contribution to the Higgs mass is up bounded by the $Z$ boson mass, 
so loop correction must be taken into account for the Higgs mass fit.
Moreover, contributions to the Higgs mass higher than the two-loop order should be considered
when the squark masses are larger than $\sim 3$ TeV.
It has been shown in \cite{1306.2318} that the three-loop contribution gives rise to 
$\sim 0.5- 3$ GeV uplifting of the Higgs mass.
For this order approximation one can either use the numerical program in \cite{1005.5709}, 
or as we choose in this paper follow the three-loop analytic formula for the Higgs mass in Ref.\cite{0701051}.

Explicitly we use the updated top quark mass $M_{t}=173.3$ GeV \cite{topmass}, 
QCD coupling structure constant $\alpha_{3}=0.1184$, 
and adopt SUSY mass parameter $\mu=200$ GeV and gluino mass $m_{\tilde{g}}=2$ TeV for the fit. 
We choose the renormalization scale $Q$ in \cite{0701051} as the the average stop mass $M_{\tilde{t}}$,
with $M^{2}_{\tilde{t}}[v]\equiv\left(m^{2}_{\tilde{t}_{1}}[v]+m^{2}_{\tilde{t}_{2}}[v]\right)/2$, 
which depends on the input values $\left(m^{2}_{\tilde{Q}_{3}},m^{2}_{\tilde{U}_{3}}\right)$
and the mixing effect $X_{t}\equiv A_{t}[v]-\mu\cot\beta$.
By using Eq.\eqref{correlation} we have, 
\begin{eqnarray}{\label{def2}}
M^{2}_{\tilde{t}}[v]=\left(y-1\right)m^{2}_{0}/2,~~~~\text{and}~~~~ X_{t}[v]\simeq A_{t}[v]=I[M]x^{1/2}m_{0}.
\end{eqnarray}

\linespread{1}\begin{figure}[!h]
 \centering%
\includegraphics[width=2.8in]{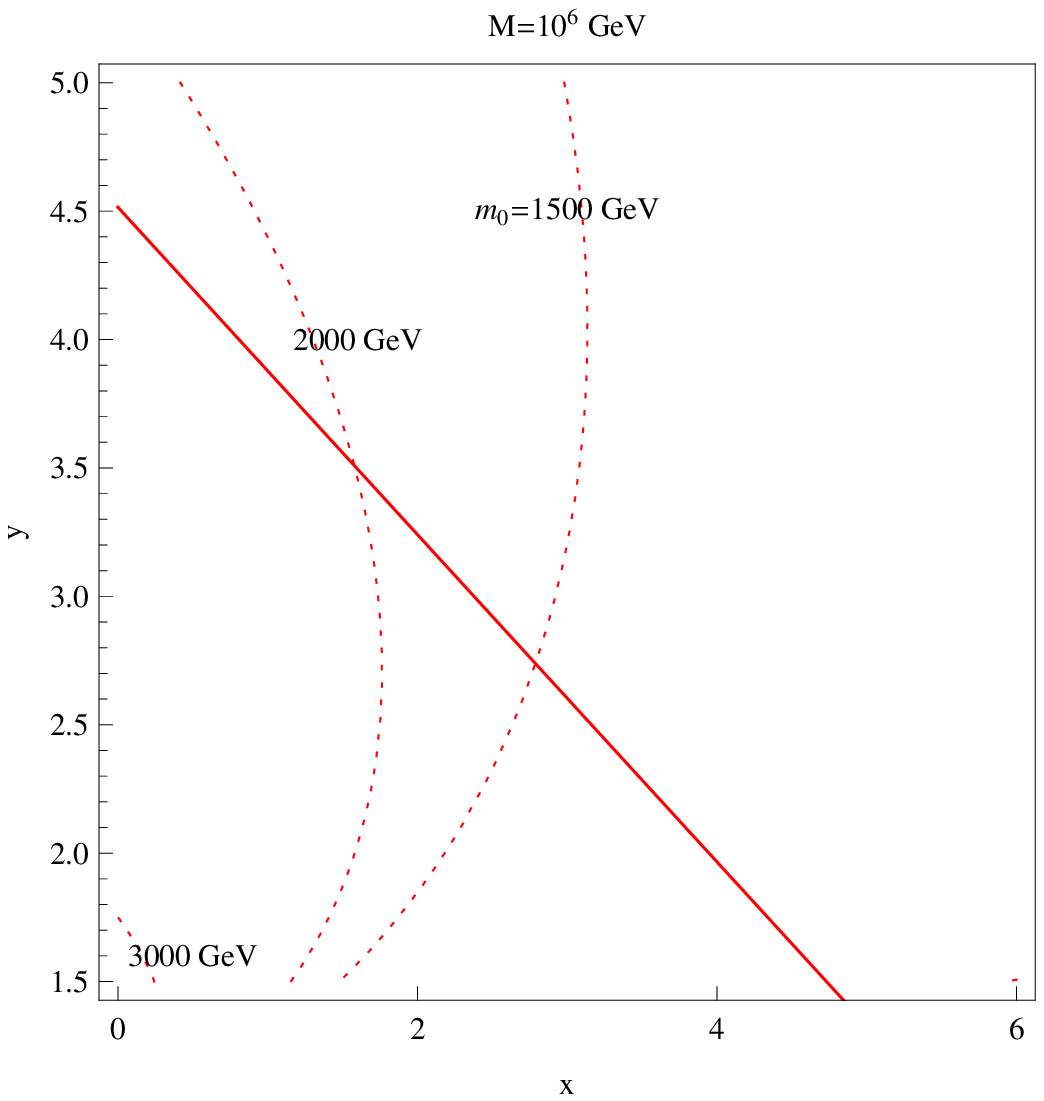}
\includegraphics[width=2.8in]{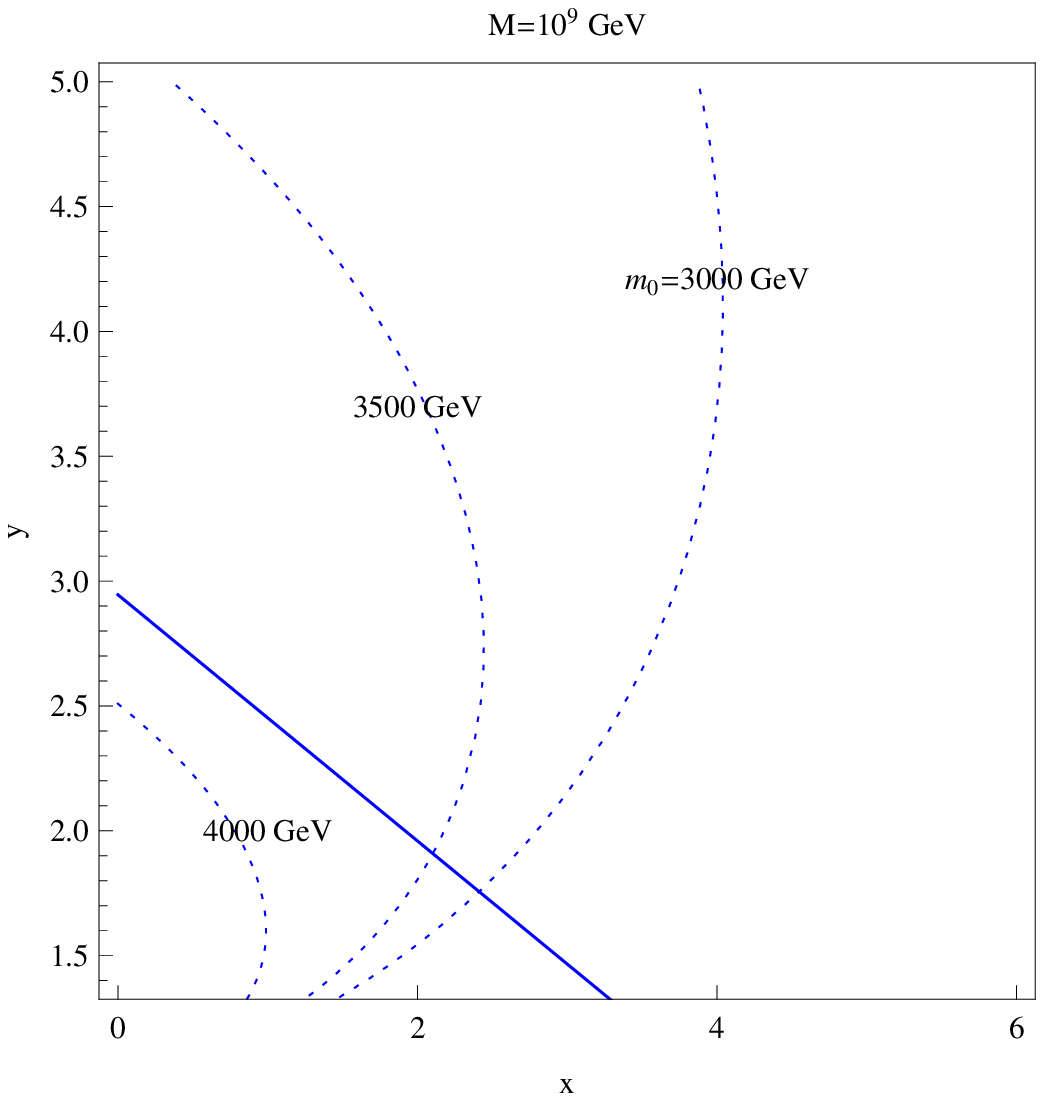}
\includegraphics[width=2.8in]{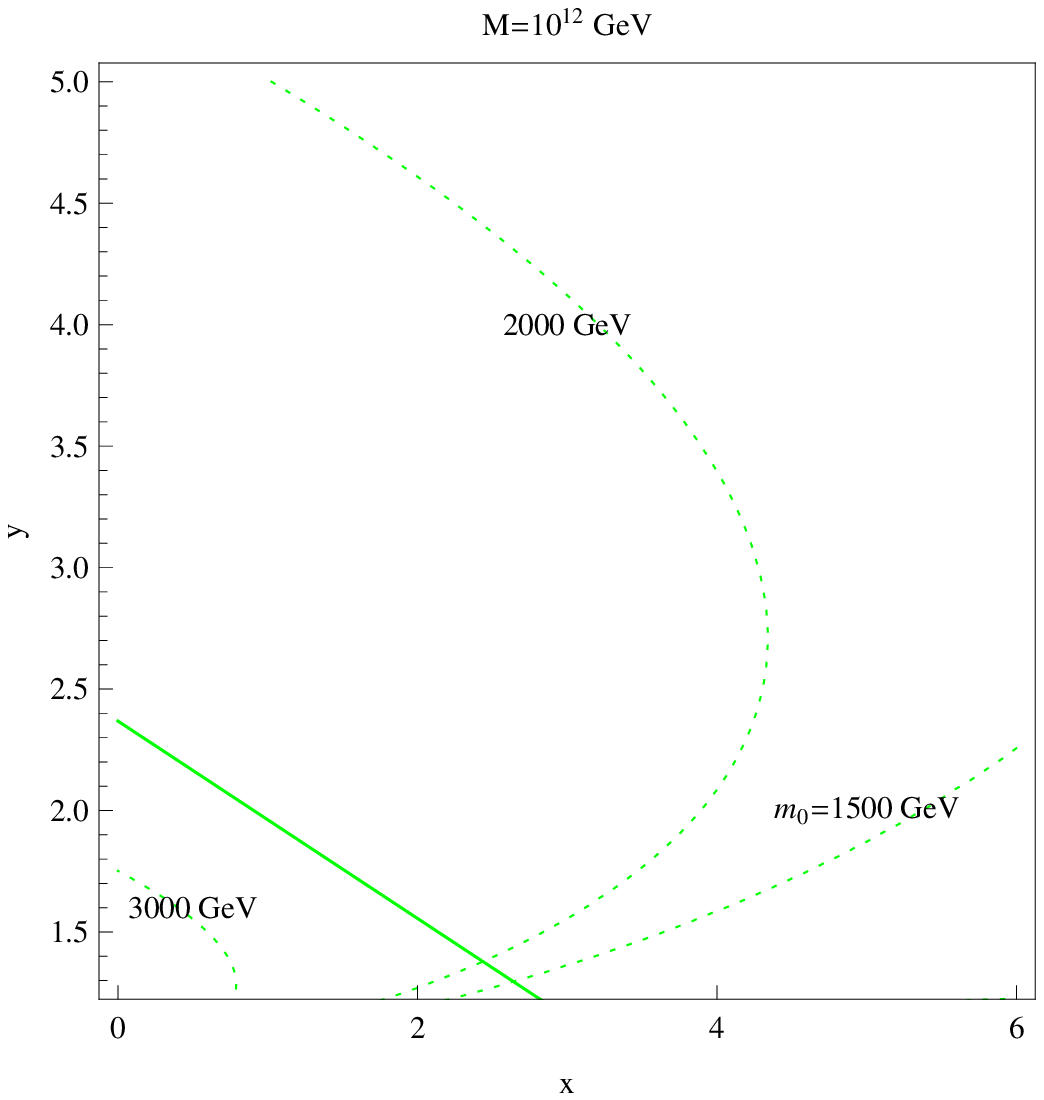}
\includegraphics[width=2.8in]{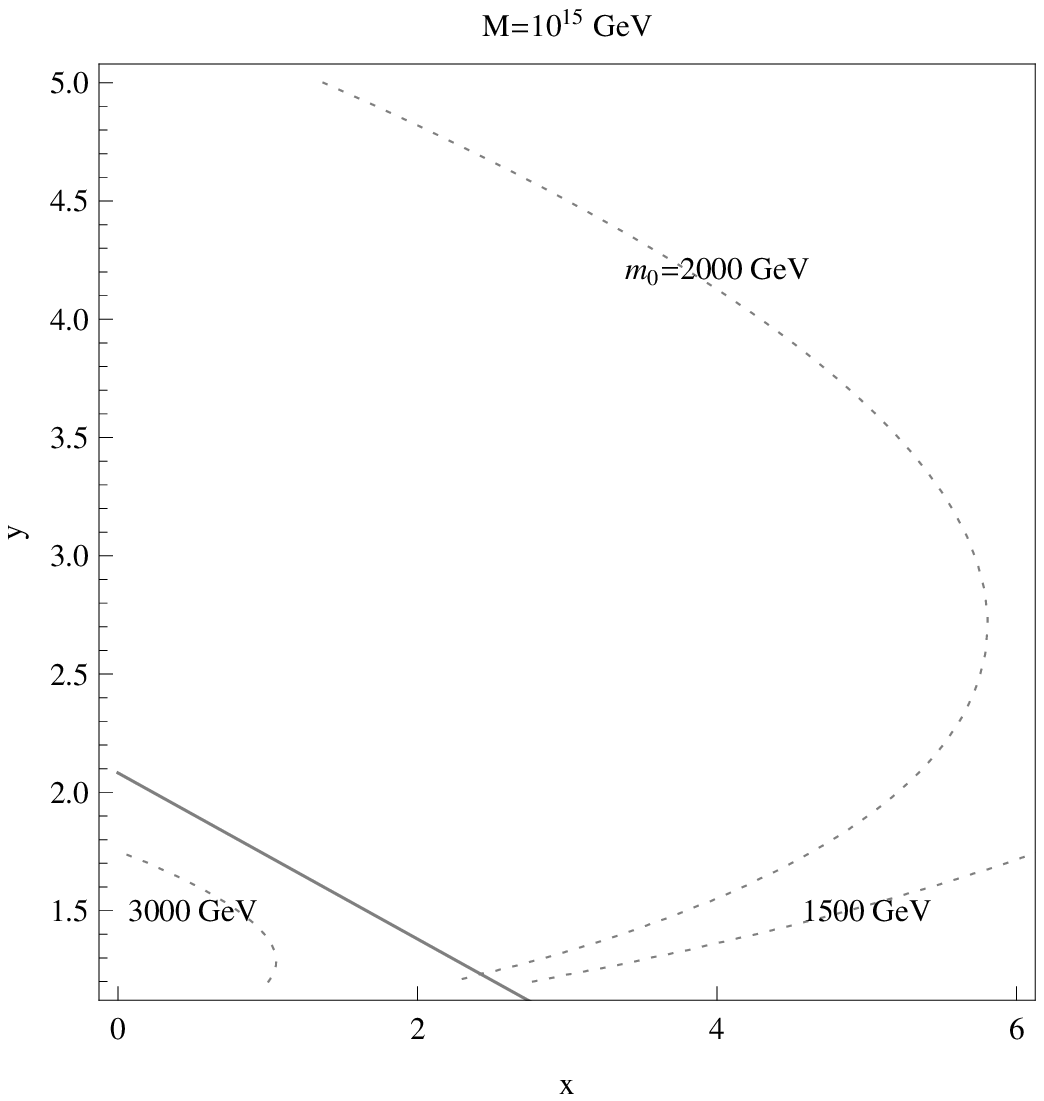}
\centering
 \caption{Contour of Higgs mass projected to the two-parameter plane of $(x,y)$ 
for $\tan\beta=20$ and $M=\{10^{6}, 10^{9}, 10^{12}, 10^{15}\}$ GeV.
The focusing lines of Fig.\ref{focuscondition} are shown simultaneously.
In each panel, we show the sensitivity of Higgs mass to the input mass parameter $m_0$.
It clearly shows that the observed Higgs mass constrains $m_0$ in the range $1$ TeV $\leq m_{0}\leq 2.5$ TeV.}
 \label{higgs}
\end{figure}

We show in Fig.\ref{higgs} the contours of three-loop Higgs mass as projected to the two-parameter plane of $(x,y)$ for $\tan\beta=20$ and $M=\{10^{6}, 10^{9}, 10^{12}, 10^{15}\}$ GeV.
Each panel corresponds to different values of $M$,
for which the focusing line in Fig.\ref{focuscondition} is shown simultaneously.
It is shown that $M_{s}[v]\simeq 3$ TeV for small mixing effect ($x\simeq 0$), 
which is consistent with the three-loop result in Ref. \cite{1306.2318} for the case of degenerate squark masses. 

\begin{figure}[!h]
 \centering%
\centering
\includegraphics[width=2.8in]{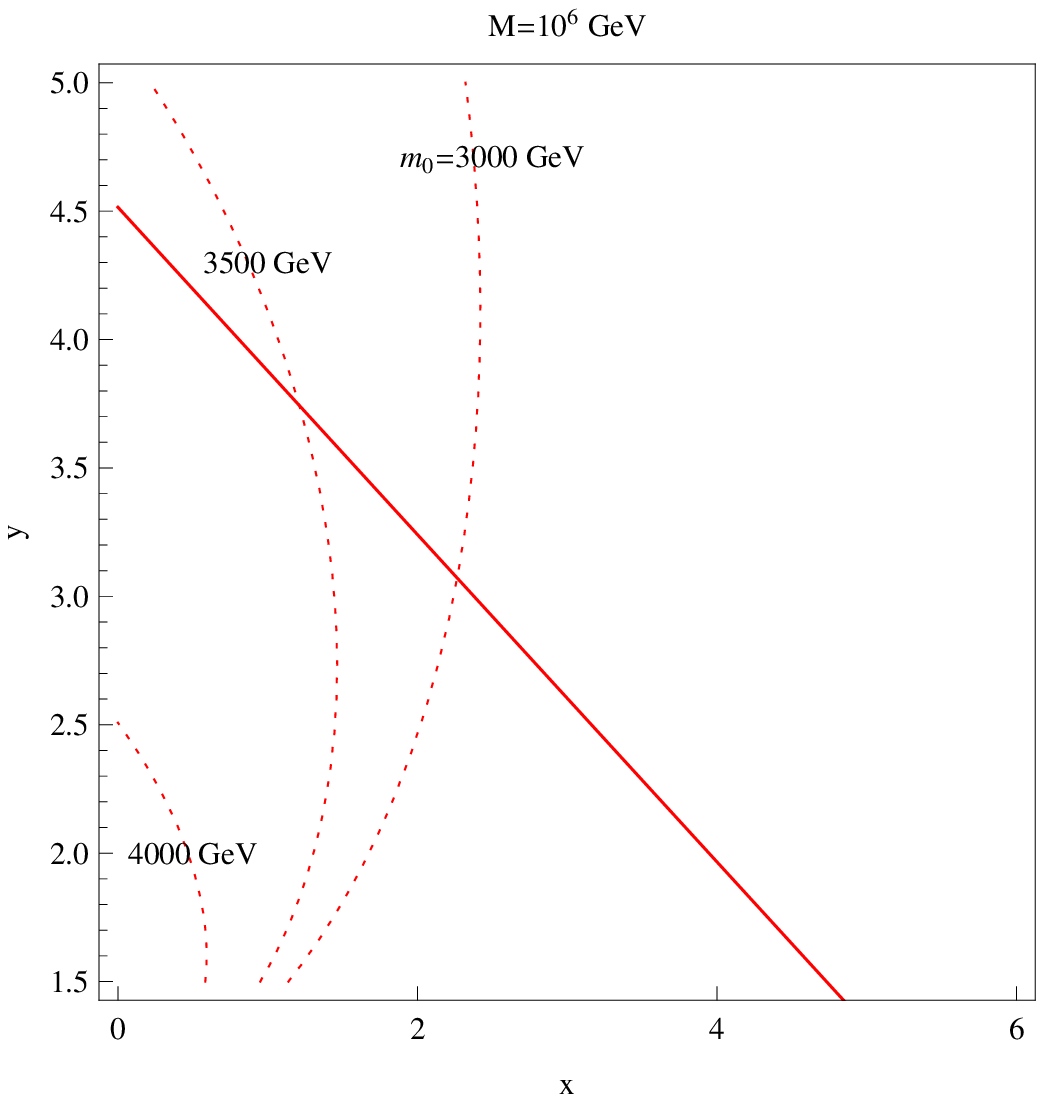}
\includegraphics[width=2.8in]{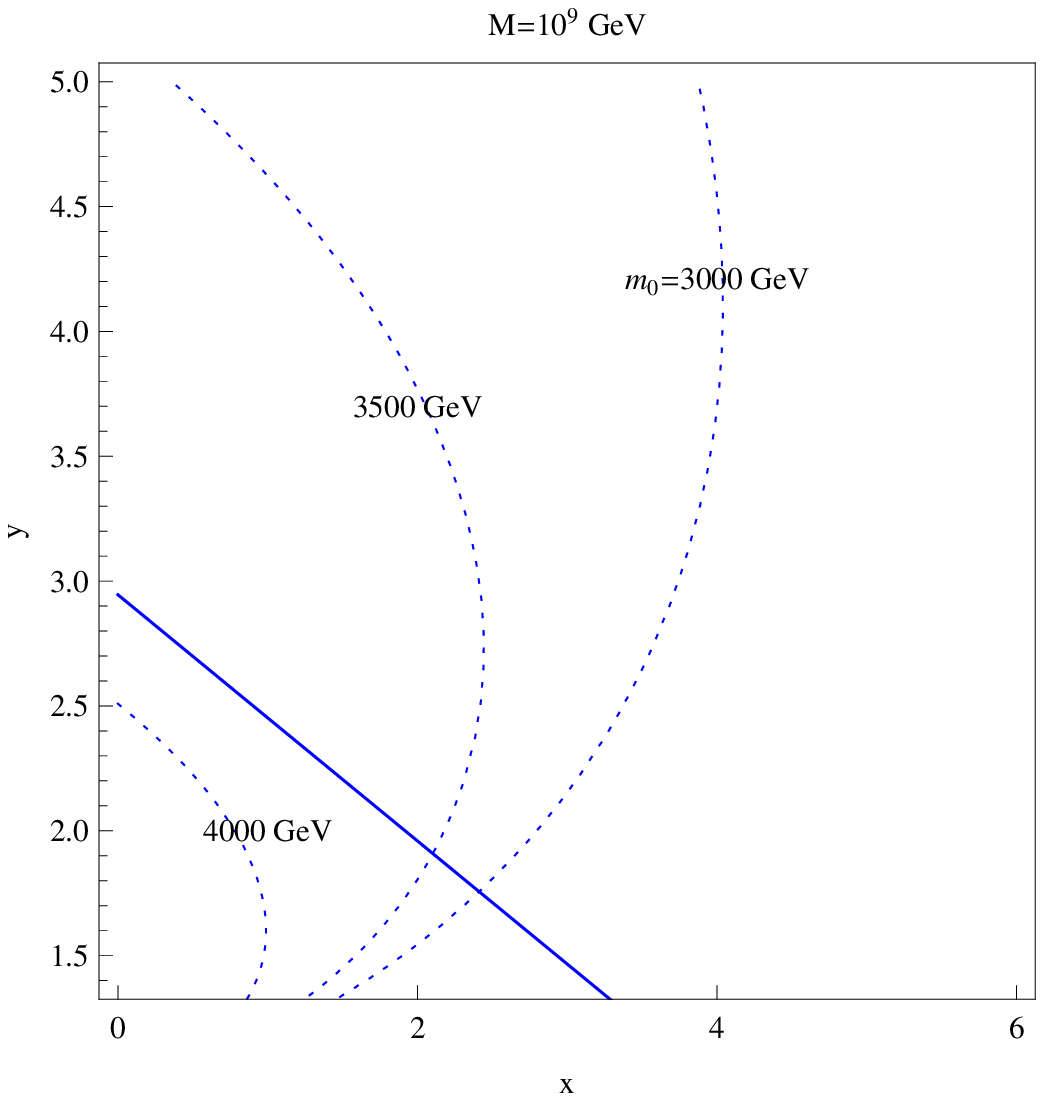}
\includegraphics[width=2.8in]{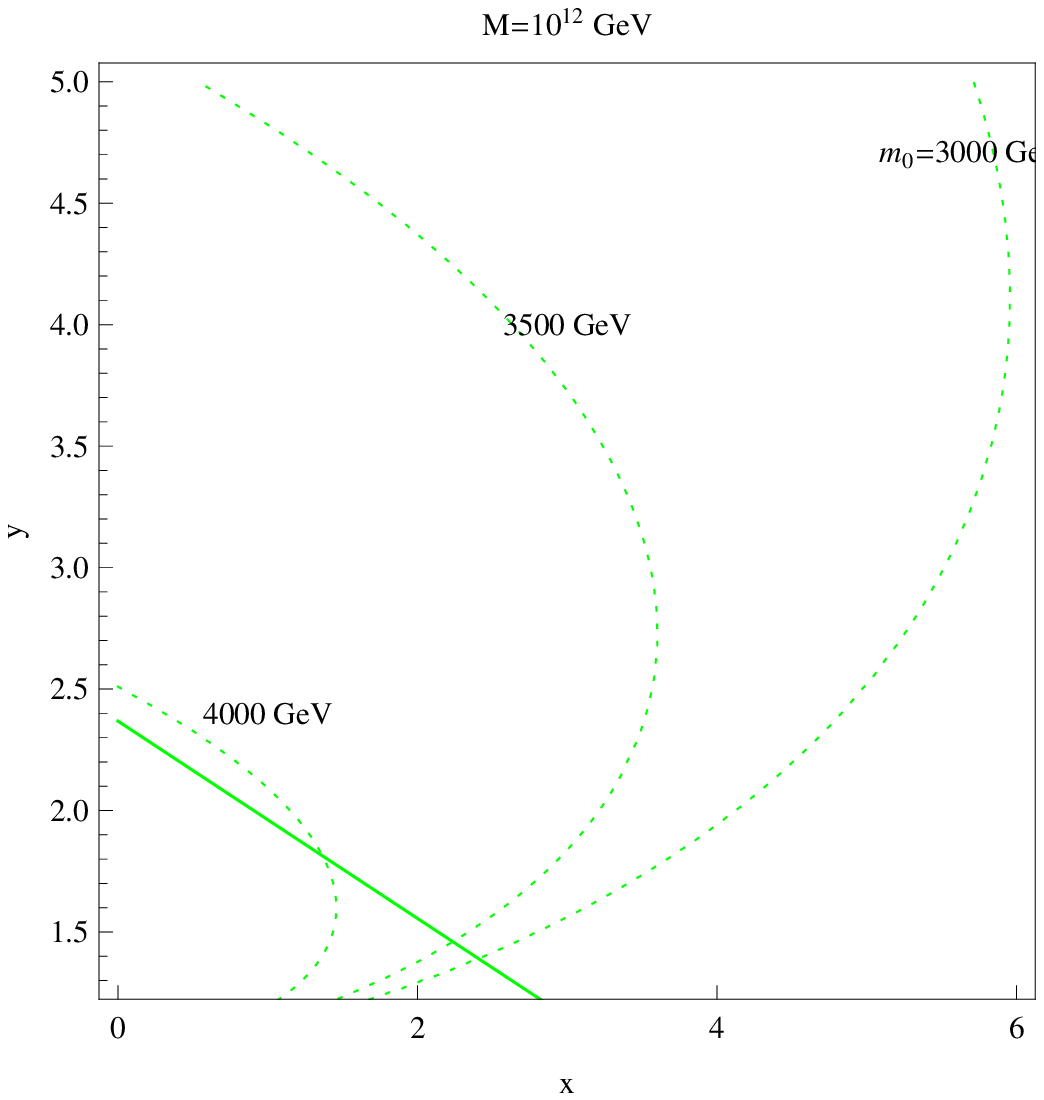}
\includegraphics[width=2.8in]{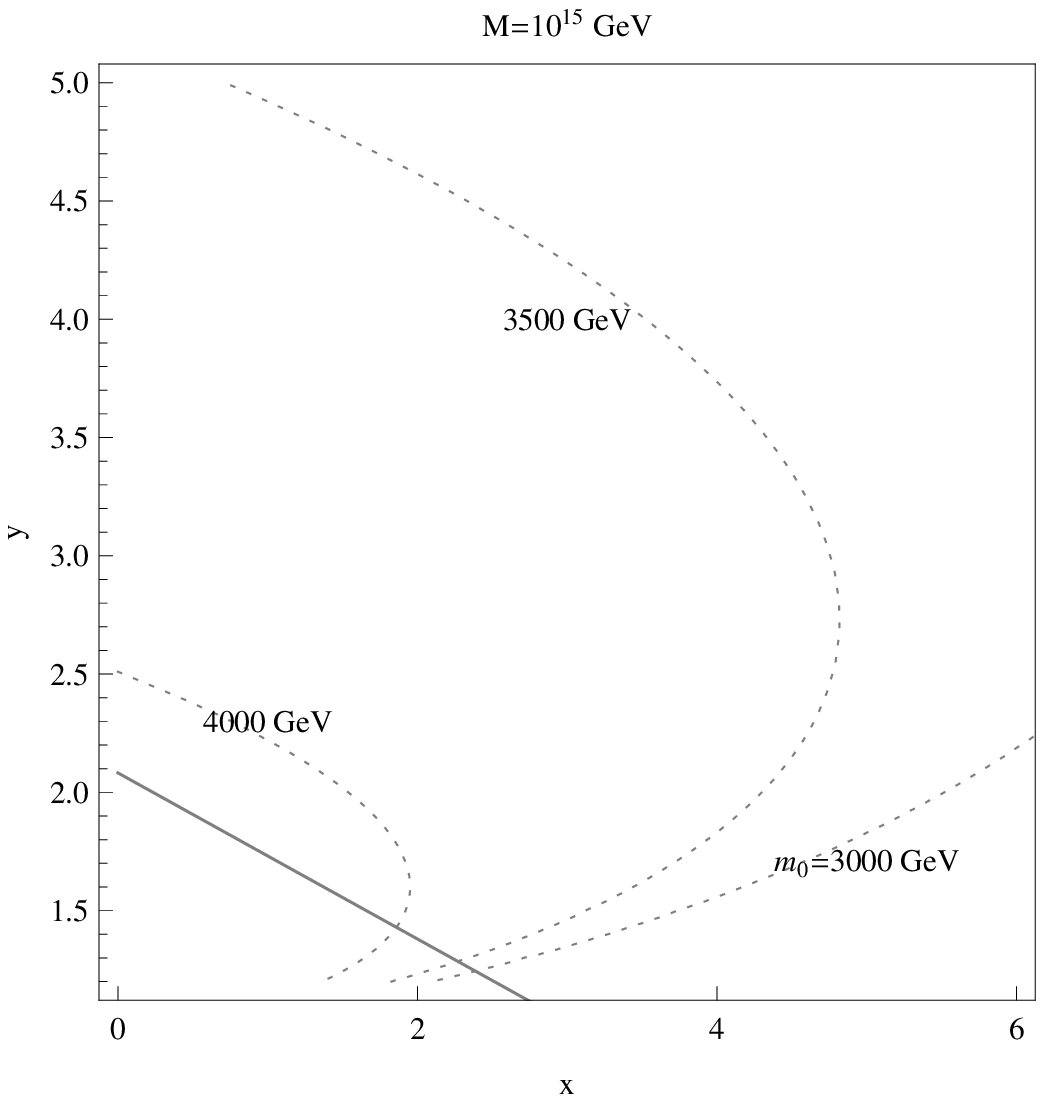}
\centering
 \caption{Same as Fig.\ref{higgs} for $\tan\beta=5$. 
In this case there is about $6$ GeV reduction in the tree-level part of Higgs mass in compared with the previous choice $\tan\beta=20$ in Fig.\ref{higgs},
so larger loop contribution to the Higgs mass is required.}
 \label{higgss}
\end{figure}

In each panel, one observes the sensitivity of Higgs mass to the input mass parameter $m_0$,
which clearly indicates that the observed Higgs mass constrains $m_0$ in the range $1.5$ TeV $\leq m_{0}\leq 3.0$ TeV. 
This range for $m_0$ is subject to the choice on parameter $\tan\beta$,
as $\tan\beta$ determines the tree-level contribution.
We show the correlation between the range and $\tan\beta$ in Fig.\ref{higgss},
which suggests that for $\tan\beta=5$ we have $3.0$ TeV $\leq m_{0}\leq 4.0$ TeV instead.  
If one takes smaller $\tan\beta$, it is expected that larger $m_{0}$ is needed. 

\linespread{1}

Here a few comments are in order, regarding the fit to the observed Higgs mass in Fig.\ref{higgs} and Fig.\ref{higgss}.

$i)$, Given the same $m_0$, compare the four contours of Higgs mass in the four panels in either Fig.\ref{higgs} or Fig.\ref{higgss}.
It is shown that the focus point value of $y$ at the crossing point decreases as $M$ approaches to the GUT scale.
The reason is partially due to the fact that $I[M]$ decreases as $M$ increases.

$ii)$, Given the same $M$, compare the contours of Higgs mass in each panel in either Fig.\ref{higgs} or Fig.\ref{higgss}.
As expected, the focus point value of $y$ ($x$) at the crossing point increases (decreases) as $m_0$ increases.
Moreover, there are more crossing points in the case with larger $M$,
In this sense the focusing may be more easily achieved in high-scale other than low-scale SUSY. 

$iii)$, For $m_{0}$ with such order of magnitude, 
the Higgs couplings of Higgs sector are similar to those in the decoupling region.
The average stop mass is above $\sim 3$ TeV for arbitrary $M$,
which has no conflict with the LHC 2013 data.
So it is probably impossible to probe so heavy stops at the second run of LHC,
and light electroweakinos will be the smoking gun of such high-scale SUSY with focus point.

In summary, the combination of Fig. \ref{higgs} and Fig. \ref{higgss} implies that 
given a focusing line referring to $M$ most of focus points are consistent with the observed Higgs mass at the LHC 
by adjusting the underlying and overall energy scale $m_0$ in the soft mass spectrum. 
Typically, $m_{0}$ is in the range of $[1.5, 2.5]$ TeV and $[3.0, 4.0]$ TeV 
for $\tan\beta=20$ and $\tan\beta= 5$, respectively.
In the next section, we will proceed to construct high-scale SUSY with focus point
which can explain the observed Higgs mass.

\section{Model Building}
In this section we consider the realization of focusing phenomenon in high-scale SUSY.
We restrict us to GM \cite{giudice} for this purpose.

\subsection{Models without $A_t$ term}
Firstly we discuss gauge mediated focus point in high-scale SUSY with small $A_t$ term.
In the context of GM, there are two important observations for our discussion.
$(1)$, the boundary value $A_{t}[M]$ disappears at the one-loop level
when there are no direct Yukawa-like couplings  between the messenger field(s) and MSSM matter field(s) 
in the superpotential.
$(2)$, the gaugino masses vanish at the same order when the mass matrix for messenger fields $\mathcal{M}$ satisfies $\det{\mathcal{M}}=\text{const}$.
$(3)$, the scalar soft masses do not disappear at this order generally. 
 
As shown in \cite{Yanagida}, 
points $(1)$-$(3)$ can be satisfied in a simple model.
In this model the messenger fields are a set of chiral and bi-fundamental 
supermultiplets $q+q'$, $\bar{q}+\bar{q'}$, $l+l'$, $\bar{l}+\bar{l'}$ 
that transforme under $SU(3)_{C} \times SU(2)_{L}\times U(1)_{Y}$ as,
\begin{eqnarray}{\label{model1}}
q, ~q'&\sim& \left(\mathbf{3}, 1, -\frac{1}{3}\right),\nonumber\\
\bar{q},~\bar{q'}&\sim& \left(\bar{\mathbf{3}}, 1, \frac{1}{3}\right),\nonumber\\
l, l'&\sim& \left(1, \mathbf{2}, \frac{1}{2}\right),\nonumber\\
\bar{l},~\bar{l'} &\sim& \left(1, \bar{\mathbf{2}}, -\frac{1}{2}\right).
\end{eqnarray}
The renormalizable superpotential consistent with SM gauge symmetry in this model reads as,
\begin{eqnarray}{\label{s1}}
W=\lambda_{2} Xl\bar{l}+\lambda_{3}Xq\bar{q}
+m_{2}\left(l'\bar{l}+l\bar{l'}\right)+m_{3}\left(q'\bar{q}+q\bar{q'}\right),
\end{eqnarray}
where SUSY-breaking sector $X=\theta^{2}F$, 
Yukawa couplings $\lambda$s and tree-level masses $m$s are assumed to be real for simplicity.

It can be easily verified that both the mass matrixes $\mathcal{M}$s for messenger vector $(q, q')$ and $(l, l')$
both satisfy $\det{\mathcal{M}}=\text{const}$.
So the gaugino masses indeed vanish at the one-loop order of $\mathcal{O}(F/M)$ as desired.
The soft scalar masses differ from those of the minimal GM,
which are given by, respectively,
\begin{eqnarray}{\label{soft}}
m^{2}_{\tilde{Q}_{3}}[M]&=& \mathcal{A}\times \left(\frac{4}{3}s_{3}\alpha^{2}_{3}[M]+\frac{3}{4}s_{2}\alpha^{2}_{2}[M]+
\frac{1}{60}s_{1}\alpha^{2}_{1}[M]\right),\nonumber\\
 m^{2}_{\tilde{U}_{3}}[M]&=&\mathcal{A}\times \left(\frac{4}{3}s_{3}\alpha^{2}_{3}[M]+
\frac{12}{45}s_{1}\alpha^{2}_{1}[M]\right),\\
m^{2}_{H_{u}}[M]&=& \mathcal{A}\times \left(\frac{3}{4}s_{2}\alpha^{2}_{2}[M]+
\frac{3}{20}s_{1}\alpha^{2}_{1}[M]\right).\nonumber
\end{eqnarray}
Here $\mathcal{A}=\frac{1}{8\pi^{2}}\frac{F^{2}}{M^{2}}$, 
which determines the overall magnitude of  soft masses above, and $s_{1, 2,3}$ are given by
\begin{eqnarray}{\label{s}}
s_{3}= \left(\lambda_{3}\frac{M}{m_{3}}\right)^{2},~~~~
s_{2}= \left(\lambda_{2}\frac{M}{m_{2}}\right)^{2},~~~~
s_{1}= \frac{3}{5}s_{2}+\frac{2}{5}s_{3}.
\end{eqnarray}
Note that the mass spectrum in Eq.\eqref{soft}  reduces to that of the minimal GM when $s_{2}=s_{3}=1$,
\begin{eqnarray}{\label{minsoft}}
m^{2}_{\tilde{Q}_{3}}[M]&=& \mathcal{A}\times \left(\frac{4}{3}\alpha^{2}_{3}[M]+\frac{3}{4}\alpha^{2}_{2}[M]+
\frac{1}{60}\alpha^{2}_{1}[M]\right),\nonumber\\
 m^{2}_{\tilde{U}_{3}}[M]&=&\mathcal{A}\times \left(\frac{4}{3}\alpha^{2}_{3}[M]+
\frac{12}{45}\alpha^{2}_{1}[M]\right),\nonumber\\
m^{2}_{H_{u}}[M]&=& \mathcal{A}\times \left(\frac{3}{4}\alpha^{2}_{2}[M]+
\frac{3}{20}\alpha^{2}_{1}[M]\right).
\end{eqnarray}

\linespread{1}\begin{figure}[!h]
\centering%
\begin{minipage}[b]{0.75\textwidth}
\centering
\includegraphics[width=4.5in]{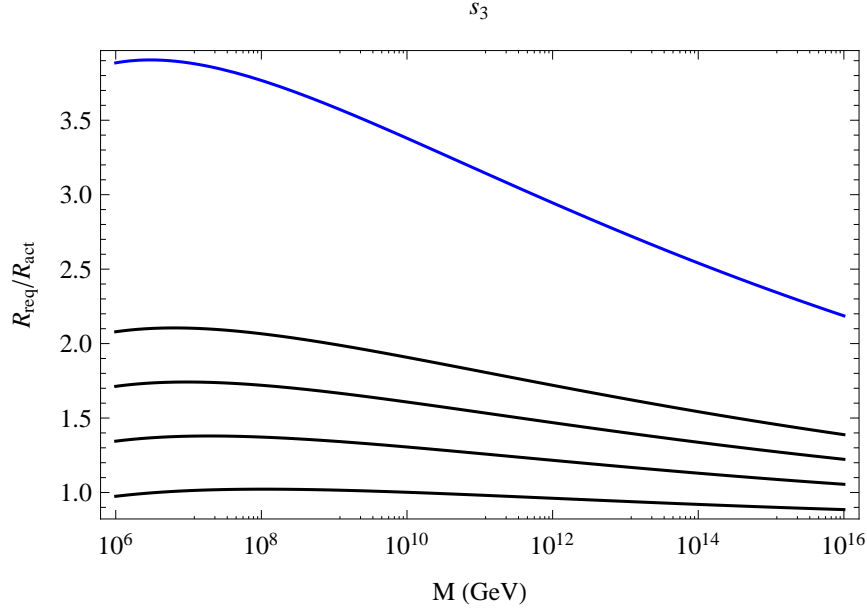}
\end{minipage}%
\caption{Ratio $\text{R}_{req}/\text{R}_{act}$ as function of $M$ 
for $s_{2}=1$ and $s_{3}=\{1,0.5,0.4,0.3,0.2\}$ from top to bottom line, respectively, 
with MSSM below scale $M$ assumed.}
\label{noAterm}
\end{figure}

\linespread{1}\begin{figure}[!h]
\centering%
\centering
\begin{minipage}[b]{0.75\textwidth}
\centering
\includegraphics[width=4.5in]{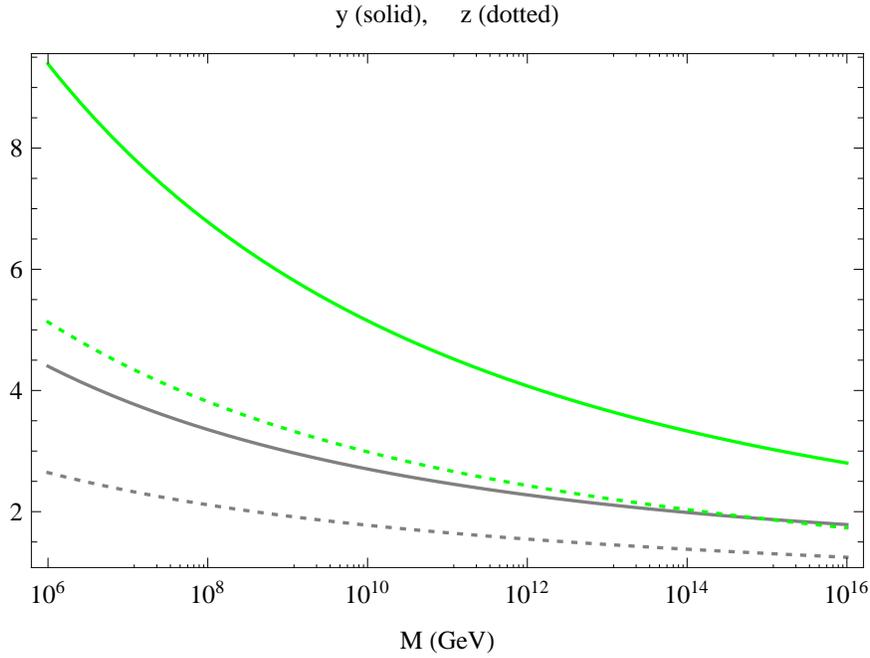}
\end{minipage}%
\caption{The value of $y$ (solid) and $z$ (dotted) for $s_{2}=1$ and $s_{3}=0.5$ $ (0.2)$  in green (gray) curve.}
\label{ss}
\end{figure}
For the case of small $A_t$ term the focusing condition in Eq.\eqref{noA} reduces to $\text{R}_{act}=\text{R}_{req}$, with 
\begin{eqnarray}{\label{ratio}}
\text{R}_{act}\equiv\frac{m^{2}_{H_{u}}[M]}{m^{2}_{\tilde{Q}_{3}}[M]+m^{2}_{\tilde{U}_{3}}[M]},~~~~~~
\text{R}_{req}=\frac{1-I[M]}{1+I[M]}.
\end{eqnarray}
In Fig.\ref{noAterm} the top line shows the ratio $\text{R}_{req}/\text{R}_{act}$ as function of $M$,
which reproduces the well known result that 
in the minimal GM $m^{2}_{H_{u}}$ is too small \cite{9910497} to satisfy the focus condition, 
which holds for the whole range of $M$.
However, the ratio changes when the soft mass spectrum deviates from the minimal GM. 
The soft mass spectrum in Eq.\eqref{soft} depends on parameter $s_{2}$ relative to $s_{3}$.
We fix $s_{2}=1$ in the following discussion,
and show that  $\text{R}_{req}/\text{R}_{act}$ is suppressed to be near unity 
by decreasing soft mass $m_{\tilde{Q}_{3}}$ and $m_{\tilde{U}_{3}}$,
i.e.,  taking smaller $s_{3}$ relative to the minimal GM case $s_{3}=1$. 
In Fig.\ref{noAterm} we show the deviation for the ratio for different values $s_{3}=\{1,0.5,0.4,0.3,0.2\}$ from top to bottom line, respectively.
The focusing condition $\text{R}_{req}/\text{R}_{act}=1$ requires $s_{3}\simeq 0.2$. 
For $s_{3}$ in the range $[0.2, 0.5]$ the consistent conditions in Eq.\eqref{range} are automatically satisfied, as shown in Fig.\ref{ss}.

\subsection{Models with $A_t$ term}
Now we consider focus point  in high-scale SUSY with large $A_t$ term.
In the model discussed in this subsection,
the messenger fields are the same as  previously studied in Eq.\eqref{model1},
except that we add a singlet of SM gauge group $S$ 
and its bi-fundamental field $\bar{S}$ into the messenger sector.
These messengers are coupled to the SUSY-breaking spurion field $X=\theta^{2}F$ through superpotential,
\begin{eqnarray}{\label{s2}}
W=X\left(l\bar{l}+q\bar{q}+\bar{S}S\right)
+m_{l}\left(l'\bar{l}+l\bar{l'}\right)+m_{q}\left(q'\bar{q}+q\bar{q'}\right).
\end{eqnarray}
We will generalize the number of messenger fields to $n$ pairs,
and simply take universal couplings between messengers and $X$ fields 
and universal masses $m_{q}\simeq m_{l}\simeq M$ for our analysis.
Similar to our previous observations,
gaugino masses vanish at the one-loop level of order $\mathcal{O}(F/M)$ in our setup. 

For our purpose we further deform the model defined in Eq.(\ref{s2}) 
by adding a direct Yukawa coupling between the lepton-like messenger $\bar{l}$ and $H_{u}$,
\begin{eqnarray}{\label{def}}
\delta W=\lambda H_{u}S\bar{l}.
\end{eqnarray}
This new superpotential can be argued to be natural by imposing hidden parity. 
As mentioned in \cite{1312.4105},  
Eq.\eqref{def} can be protected in terms of either imposing some hidden matter parity other than $R$ parity
or global $U(1)$ symmetry.
If so, Eq.\eqref{def} ensures an one-loop $A_t$ term of order $\mathcal{O}(F/M)$ at the messenger scale.

In this model scalar soft mass spectrum includes the contribution from Eq.(\ref{s2}) and the new one induced 
by Yukawa coupling $\lambda$ in Eq.\eqref{def}.
For the case of small SUSY breaking ($F< M^{2}$) the former  is given by,
\begin{eqnarray}{\label{softold}}
m^{2}_{\tilde{Q}_{3}}[M]&=& \mathcal{A}' \cdot \left(\frac{4}{3}\alpha^{2}_{3}[M]+\frac{3}{4}\alpha^{2}_{2}[M]+
\frac{1}{60}\alpha^{2}_{1}[M]\right),\nonumber\\
 m^{2}_{\tilde{U}_{3}}[M]&=&\mathcal{A}' \cdot \left(\frac{4}{3}\alpha^{2}_{3}[M]+
\frac{12}{45}\alpha^{2}_{1}[M]\right),\nonumber\\
m^{2}_{H_{u}}[M]&=& \mathcal{A}' \cdot \left(\frac{3}{4}\alpha^{2}_{2}[M]+
\frac{3}{20}\alpha^{2}_{1}[M]\right).
\end{eqnarray} 
with $\mathcal{A}'=\frac{n}{8\pi^{2}}\frac{F^{2}}{M^{2}}$.
While the later one reads as,
\begin{eqnarray}{\label{softnew}}
\delta m^{2}_{\tilde{Q}_{3}}[M]&\simeq&\frac{1}{2}\mathcal{A}' \cdot \left(-\frac{n}{2}\alpha_{t}[M]\alpha_{\lambda}[M]\right),\nonumber\\
 \delta m^{2}_{\tilde{U}_{3}}[M]&\simeq&\frac{1}{2}\mathcal{A}' \cdot\left(-n\alpha_{t}[M]\alpha_{\lambda}[M]\right),\nonumber\\
\delta m^{2}_{H_{u}}[M]&\simeq&\frac{1}{2}\mathcal{A}' \cdot \left\{\frac{1}{2} n(n+3)\alpha^{2}_{\lambda}[M]-n\alpha_{\lambda}[M]\left(\frac{3}{10}\alpha_{1}[M]+\frac{3}{2}\alpha_{2}[M]\right)\right\},\nonumber\\
\delta A^{2}_{t}[M]&\simeq&\frac{1}{2}\mathcal{A}' \cdot \left(\frac{1}{2}n^{2}\alpha^{2}_{\lambda}[M]\right).
\end{eqnarray}

Here a few comments are in order, regarding Eq.(\ref{softnew}).
The contribution to soft scalar mass spectrum 
due to Yukawa-type interaction has been previously studied in \cite{1206.4086, 9706540, 0112190}.
However, they can not be directly applied to our case.
Because the SUSY-breaking spurion superfield $X$ differs from the conventional situation in which $X_{\text{stand}}=\left<X\right>+F\theta^{2}$, with $\left<X\right>\neq0$.
We present the derivation of Eq.(\ref{softnew}) in appendix A.
Focusing phenomenon in this model has been discussed in \cite{1312.4105} for intermediate scale $M\sim 10^{8}$ GeV. Here, our discussions are more general, with scale $M$ unfixed.

\linespread{1}\begin{figure}[!h]
\centering%
\centering
\begin{minipage}[b]{0.75\textwidth}
\centering
\includegraphics[width=4.5in]{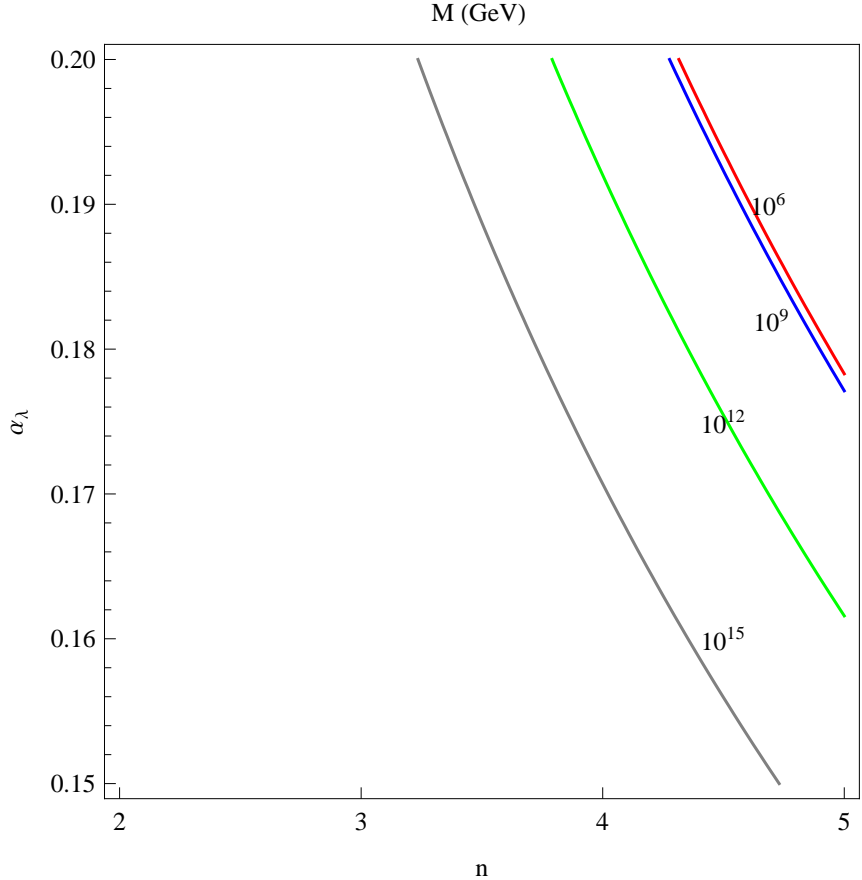}
\end{minipage}%
\caption{Solution to the focusing condition in Eq.\eqref{condition} projected to two-parameter plane of $(n, \alpha_{\lambda_{S}})$ 
for $M=\{10^{6}, 10^{9}, 10^{12}, 10^{15}\}$ GeV.
It has been verified that each solution satisfies the consistent conditions in Eq.\eqref{range}.}
\label{na}
\end{figure}

In this model the input parameters related to focusing conditions Eq.\eqref{condition} and Eq.\eqref{range}
at high energy scale $M$ are composed of the number of messenger pairs $n$, 
Yukawa coupling constant $\alpha_{\lambda}[M]$ and scale $M$.
The other two parameters $\mathcal{A}'$ and $\tan\beta$ are related to the observed Higgs mass at the weak scale.
Note that requiring grand unification of SM gauge couplings leads to $n\leq 5$,
and requiring the absence of Landau pole leads to an upper bound on $\alpha_{\lambda}[v]$,
although the later of which can be relaxed by either embedding the model into strong or conformal dynamics.
We discuss these two parameters in the range below,
\begin{eqnarray}{\label{p}}
1\leq n\leq 5, ~~~~~\text{and} ~~~~\alpha_{\lambda}[v]\leq 1/4\pi.
\end{eqnarray}
In Fig.\ref{na} we show the solution to focusing condition Eq.\eqref{condition} 
in the two-parameter plane of $(n, \alpha_{\lambda})$ for different values of $M$.
We have verified that each solution satisfies the consistent conditions in Eq.\eqref{range}.
Fig.\ref{na} clearly indicates that focusing may happen for small $n\simeq 3$ and small $\alpha_{\lambda}[M]\sim 0.2$ when $M$ is near the GUT scale.
Conversely, large $n\simeq 5$ and large $\alpha_{\lambda}[M]\simeq 0.18$ are  required when $M$ is small $\sim 10^{6}$ GeV.
The RG effect on $\alpha_{\lambda}$ actually excludes such large $\alpha_{\lambda}[v]$ in perturbative theory.
For example, it requires $M> 10^{9}$ GeV for the representative value $\alpha_{\lambda}[v]= 1/4\pi$.

In summary,  focus point high scale SUSY can be realized in non-minimal GM.
But only large $M$ is allowed for the case with large $A_t$ term.
In the next subsection, we address the problem of small gaugino mass in general focus point SUSY,
which can be reconciled with the LHC 2013 data in terms of reasonable modification to the original messenger sector.

\subsection{Gaugino Mass}
The LHC 2013 data \cite{ATLAS2013, CMS2013} has reported gluino mass bound about $\sim 1.3$ TeV.
This mass bound is not satisfied in the focus point SUSY discussed so far. 
For example, the gluino mass relative to $m_{H_{u}}$ at scale $M$ in the case with large $A_t$ discussed in IV.B is given by,
\begin{eqnarray}{\label{gluino}}
\frac{m_{\tilde{g}_{3}}[M]}{m_{0}} \simeq 4\sqrt{2}\pi^{2}\cdot\frac{s^{5/2}_{3}\alpha_{3}[M]}{\left(\frac{3}{4}s_{2}\alpha^{2}_{2}[M]+
\frac{3}{20}s_{1}\alpha^{2}_{1}[M]\right)^{3/2}} \frac{m^{2}_{0}}{M^{2}}\cdot\mathcal{F}\left(\frac{-s_{3}+\sqrt{s^{2}_{3}+4}}{2},\frac{-1+\sqrt{5}}{2}\right)\nonumber\\
\end{eqnarray}
where function $\mathcal{F}(a,b)$ is defined in \cite{Yanagida}.
Together with the constraint on $m_{0}$ due to observed Higgs mass as shown in the section III,
one can estimate the boundary value of $m_{\tilde{g}_{3}}[M]$.
Eq.\eqref{gluino} implies that the gluino mass is far below the present lower bound.

Some reasonable modification should be taken into account 
in order to complete the discussions about model building.
Now we re-examine the smallness of gaugino masses,
which attributes to the fact that $\det{\mathcal{M}}=\text{const}$ 
and consequently $m_{\tilde{g}_{r}}\sim \frac{\alpha_{r}}{4\pi}F\partial \ln\det{\mathcal{M}}/\partial \ln X\sim 0$.
When we employ small tree-level mass terms for some of messengers such as
\footnote{Note that these mass terms are consistent with gauge symmetries and matter parity of messenger sector.}
\begin{eqnarray}{\label{mass}}
\delta W= m'\bar{q'}q' + m' \bar{l'}l',
\end{eqnarray}
it will lead to the replacement in $\det{\mathcal{M}}$ for quark and lepton messengers \cite{0612139},
\begin{eqnarray}{\label{mod}}
\mathcal{M}=\left(
\begin{array}{ll}
X & M \\
M & 0 
\end{array}\right)\rightarrow
\left(
\begin{array}{ll}
X & M \\
M & m'
\end{array}\right).
\end{eqnarray}
If so, the correction to soft scalar mass spectrum and gaugino masses 
is of order $\mathcal{O}(m'^{4}/M^{4})$ and
\begin{eqnarray}{\label{g2}}
m_{\tilde{g}_{i}}\simeq n \cdot\frac{\alpha_{i}}{4\pi}\cdot\frac{F}{M}\cdot\frac{m'}{M},
\end{eqnarray}
, respectively.
Provided 
\begin{eqnarray}{\label{con}}
 \frac{F^{2}}{M^{4}}<<\frac{m'}{M}<<1,
\end{eqnarray}
the former correction can be very weak so that the focusing still holds, 
but the later one can be large enough to reconcile with the LHC gaugino mass bound. 
For example,  we choose $m'\simeq 0.06 M\simeq10^{8}$ GeV.
For $m_{0}\sim 3$ TeV we have $\sqrt{F}\sim 6\times 10^{7} $ GeV, 
and further $m_{\tilde{g}_{3}}\simeq 2$ TeV from Eq.(\ref{g2}).
In this model both the bino and wino masses are around $\sim1$ TeV,
so they are the main target at the 14-TeV LHC.
The discussions about model building in high-scale SUSY with focus point are therefore completed.

\section{Conclusion}
In this paper we have explored  focus point in the high-scale SUSY which has weak-scale electroweakino masses.
We have derived conditions for the focusing phenomenon in general.
We have analyzed in detail the prediction for the Higgs mass in such focus point SUSY.
The observed Higgs mass at the LHC  requires the  input value $m^{2}_{H_{u}}$ of order $\sim 2$ TeV
and $\sim 3$ TeV for $\tan\beta=\{20,5\}$, respectively.
We also address the model building of focus point SUSY by employing non-minimal GM.
The main results include 
$(a)$ for the case with small $A_t$ term, focus point is allowed for a wide range of $M$; 
$(b)$ for the case with large $A_t$ term, focus point is only permitted in high-scale SUSY.

One may worry about the stability of focus point discussed so far by following two facts.
One fact is that the soft scalar mass spectrum is directly related 
and thus sensitive to the underlying mass scales $F$ and $M$.
The other one is that the focusing phenomenon imposes constraint on these soft masses as shown in Eq.(\ref{mH}). 
So, the significant reduction due to focus point seems to be spoiled for a small change of either $F$ or $M$.
However, this is not true.
Because only the overall magnitude of soft mass spectrum but not their relative ratios is determined by $F/M$.
As shown in Eq.(\ref{condition}) the focusing condition Eq.(\ref{mH}) is actually not dependent on either $F$ or $M$,
and only dependent on $M$ indirectly through RG effect.
Once we fix $M$, for example in some GUT-scale SUSY models, 
the focus point induced by the accidental  cancellation,
is actually stable against to high mass scale.

\begin{acknowledgments}
This work is supported in part by the Natural Science Foundation of China under grant No.11247031 and 11405015.
\end{acknowledgments}

\appendix
\section{Soft Mass Spectrum}
The calculation of soft mass spectrum in GM can be performed in a few ways.
Among them evaluating Feynman diagrams is the most direct approach \cite{9608224}.
However, the number of Feynman diagrams for calculating these soft scalar mass is usually large,
which limits its application broadly.
In \cite{9706540} a new technique is proposed, 
where the authors suggest analytical continuation into superspace $M\rightarrow  X_{\text{stand}}=M+F\theta^{2}$ in the renormalized wave function, 
where $M$ and $F$ refers to the messenger mass scale and the strength of SUSY breaking, respectively.
Nevertheless,  this method is only valid for the standard spurion superfield $X_{\text{stand}}$ 
and for the case of small SUSY breaking.

Therefore, for non-standard spurion field $X=F\theta^{2}$ as studied in this paper,
we have to evaluate Feynman diagrams directly.
By following the results  \cite{9608224} for standard spurion $X_{\text{stand}}$ 
in \cite{9708377} the authors have derived the soft scalar mass spectrum in the model  defined in Eq.(\ref{s2}),
where it is crucial to establish the connection between the mass eigenstates of messenger fields in these two cases.
Following this insight, we derive the correction due to Yukawa superpotential deformation Eq.(\ref{def}) 
in terms of results in the standard GM with Yukawa superpotential deformation Eq.(\ref{def}).

In terms of Eq.(\ref{s2}) one obtains the mass matrixes
\begin{eqnarray}{\label{a3}}
\left(%
\begin{array}{cc}
  \mid l\mid, & \mid \bar{l}\mid \\
\end{array}%
\right)\left(%
\begin{array}{cc}
  M^{2} & F \\
  F^{*}& M^{2}\\
\end{array}%
\right)\left(%
\begin{array}{cc}
 \mid l\mid \\
 \mid \bar{l}\mid\\
\end{array}%
\right)
\end{eqnarray}
and 
\begin{eqnarray}{\label{a4}}
\left(%
\begin{array}{cc}
  \mid l'\mid, & \mid \bar{l'}\mid \\
\end{array}%
\right)\left(%
\begin{array}{cc}
  M^{2} & 0 \\
  0 & M^{2}\\
\end{array}%
\right)\left(%
\begin{array}{cc}
 \mid l'\mid \\
 \mid \bar{l'}\mid\\
\end{array}%
\right)
\end{eqnarray}
for  lepton messenger scalars, and 
\begin{eqnarray}{\label{a5}}
\left(%
\begin{array}{cc}
  \psi_{\bar{l}}, &  \psi_{\bar{l'}} \\
\end{array}%
\right)\left(%
\begin{array}{cc}
  0 & M \\
  M & 0\\
\end{array}%
\right)\left(%
\begin{array}{cc}
\psi_{l} \\
\psi_{l'}\\
\end{array}%
\right)
\end{eqnarray}
for  lepton messenger fermions.
The scalar mass matrix in Eq.(\ref{a3}) is the same as minimal GM,
while the scalar mass matrix in Eq.(\ref{a4}) is diagonal.
So only the fermion mass matrix $\mathcal{M}^{(\psi)}$ in Eq.(\ref{a5}) needs to be diagonalized.
Similar to \cite{9708377} we assume an unitary matrix $V(\theta)$ to achieve this,
with $\theta$ an angle.
In principle, we need two independent unitary matrixes for more general $\mathcal{M}^{(\psi)}$ other than that in Eq.(\ref{a5}).
The two eigenstate masses for fermion messengers are equal to $M$.
Consequently, we build the connection between gauge and mass eigenstates as,
\begin{eqnarray}{\label{a6}}
\left(%
\begin{array}{cc}
\psi_{l} \\
\psi_{l'}\\
\end{array}%
\right)\rightarrow
\left(%
\begin{array}{cc}
  1 & 0 \\
  0 & -1\\
\end{array}%
\right)
\left(%
\begin{array}{cc}
  \cos\theta & -\sin\theta \\
 \sin\theta& \cos\theta\\
\end{array}%
\right)\left(%
\begin{array}{cc}
\psi_{l} \\
\psi_{l'}\\
\end{array}%
\right)
\end{eqnarray}
and,
\begin{eqnarray}{\label{a7}}
\left(%
\begin{array}{cc}
\psi_{\bar{l}} \\
\psi_{\bar{l'}}\\
\end{array}%
\right)\rightarrow
\left(%
\begin{array}{cc}
  \cos\theta & -\sin\theta \\
 \sin\theta& \cos\theta\\
\end{array}%
\right)\left(%
\begin{array}{cc}
\psi_{\bar{l}} \\
\psi_{\bar{l'}}\\
\end{array}%
\right)
\end{eqnarray}

The correction to soft scalar mass spectrum due to the Yukawa superpotential deformation 
to the standard GM has been previously studied in \cite{1206.4086, 9706540, 0112190}.
We find that the the messenger sector is divided into two independent subsectors $(l', \bar{l'})$ and $(l, \bar{l})$,
and masses for messenger scalars and fermions in the subsector $(l', \bar{l'})$ are degenerate at $M$.
So the correction to soft mass spectrum only attributes to subsector $(l, \bar{l})$, 
which reads as,
\begin{eqnarray}{\label{a6}}
\delta m^{2}_{\tilde{Q}_{3}}[M]&\simeq&\mathcal{A}' \cdot \cos^{2}\theta \cdot \left(-\frac{1}{2}d_{H}\alpha_{t}[M]\alpha_{\lambda}[M]\right),\nonumber\\
 \delta m^{2}_{\tilde{U}_{3}}[M]&\simeq&\mathcal{A}' \cdot \cos^{2}\theta \cdot \left(-d_{H}\alpha_{t}[M]\alpha_{\lambda}[M]\right),\\
\delta m^{2}_{H_{u}}[M]&\simeq&\mathcal{A}' \cdot \cos^{2}\theta \cdot  \left(\frac{1}{2} d_{H}(d_{H}+3)\alpha^{2}_{\lambda}[M]-d_{H}\alpha_{\lambda}[M]\sum_{r=1}^{r=3}C^{r}\alpha_{r}[M]\right),\nonumber\\
\delta A^{2}_{t}[M]&\simeq&\mathcal{A}' \cdot \cos^{2}\theta \cdot  \left(\frac{1}{2}d^{2}_{H}\alpha^{2}_{\lambda}[M]\right),\nonumber
\end{eqnarray}
with $\mathcal{A}'=\frac{n}{8\pi^{2}}\frac{F^{2}}{M^{2}}$,
Here the structure constants  are defined as,
\begin{eqnarray}{\label{alpha}}
\alpha_{r}[M]\equiv\frac{g^{2}_{r}[M]}{4\pi},~~~~~~~~
\alpha_{t}[M]\equiv\frac{y_{t}^{2}[M]}{4\pi},~~~~~~~~
\alpha_{\lambda_{S}}[M]\equiv\frac{\lambda_{S}^{2}[M]}{4\pi},
\end{eqnarray}
with $g_r$ ($r=1,2,3$) refer to the SM gauge couplings.
For the model under study we have $\tan\theta =1$ and $d_{H}=n$, $n$ being number of messenger pairs.
$C^{r} =C^{r}_{H_{u}}+C^{r}_{i} +C^{r}_{j}$ is the sum of quadratic Casimirs of the fields which participate in the Higgs-messenger-messenger Yukawa coupling,  
with $i, j$ referring to messenger fields involved.
In our case, $C^{r}=\left(\frac{3}{10}, \frac{3}{2}, 0\right)$.

\linespread{1}
\end{document}